\documentstyle[12pt,epsfig]{article}

\begin{document}
\begin {center}
{\bf {\Large
Nuclear effects on the $\rho$ meson produced in the inclusive
photonuclear reaction
} }
\end {center}
%\smallskip
%\medskip
\begin {center}
Swapan Das \footnote {email: swapand@barc.gov.in} \\
{\it Nuclear Physics Division,
Bhabha Atomic Research Centre,  \\
Trombay, Mumbai: 400085, India \\
Homi Bhabha National Institute, Anushakti Nagar,
Mumbai-400094, India }
\end {center}
%\medskip

\begin {abstract}

The hadronic properties of the $\rho$ meson produced in the inclusive
photonuclear reaction have been investigated. The elementary reaction
occurring in the nucleus is assumed  as $\gamma N \to \rho^0 N$.
The
$\rho$ meson, while propagating through the nucleus, interacts with the
nuclear particles, and therefore, the properties of the $\rho$ meson can be
modified because of this interaction.
Being
a short-lived particle, the $\rho$ meson decays to various elementary
particles, such as, $e^+e^-$, $\pi^+\pi^-$, .... etc. The $e^+e^-$ invariant
mass, i.e., the $\rho$ meson mass, distribution spectra have been
calculated to extract the information about the parameters, viz., mass and
width, of the $\rho$ meson in the nucleus. The calculated results have been
compared with the data reported from Jefferson Laboratory.

\end {abstract}

%Keywords:

%Photonuclear reaction, $\rho$ meson modification 

%\smallskip
%PACS number(s): xxx, xxx

\section{Introduction}

The modification of the hadronic properties, i.e., mass and width, of vector
mesons (e.g., $\rho$, $\omega$ and $\phi$ mesons) in a nucleus is an
important aspect in the nuclear physics, particularly in context to restore
the chiral symmetry of those mesons \cite{brown}.
The
calculated results show the modification of the vector meson in a normal
nucleus is significant. The shape of the $\rho$ meson mass distribution
spectrum in the pion nucleus reaction is deferred from that in the free space
\cite{golu, weid}.
The
modification of the hadronic parameters is shown to occur for the $\rho$ and
$\omega$ mesons produced in the photonuclear reaction \cite{effen, muhl}. The
in-medium properties of the $\phi$  meson produced in the nuclear reactions
are discussed in Ref.~\cite{cabr}.

The medium modification of the vector mesons has been explored by measuring
the invariant mass distribution spectrum of their decay products.
In
the heavy-ion collision experiments, the broadening of the $\rho$ meson width in
the dielectron and dimuon invariant mass distribution spectra are reported
by CERES \cite{adam} and NA60 \cite{arnal} collaborations respectively.
The
modification of the $\rho$, $\omega$ and $\phi$ mesons in the proton-nucleus
collision has been investigated by the KEK-PS \cite{naruki} and ANKE
collaborations \cite{poly}.
The
previous collaboration has demonstrated the modification in the spectral
shape of these mesons due to excess mass on the lower side of the $\omega$
and $\phi$ mesons' peak in the dilepton $e^+e^-$ invariant mass distribution
spectra, where as the latter collaboration has revealed large spreading in
the width of the $\phi$ meson in the dikaon $K^+K^-$ invariant mass
distribution spectrum.
The
CB-ELSA collaboration \cite{nanov, thiel} could not find the mass-shift of
the $\omega$ meson, produced in the photonuclear reaction, in the measured
$\pi^0 \gamma$ invariant mass distribution spectrum. This observation is
supported by the calculated results \cite{dasL}. However, the above
collaboration has reported the enhancement in the width of the $\omega$ meson
\cite{thiel, kotu}.
It
should be mentioned that the KEK data (expressing the vector meson mass
modification) suffer from the background subtraction which lead to erroneous
conclusions, as pointed out by the CLAS collaboration \cite{nass, wood}.
This
collaboration carried out the systematic investigation of the in-medium
properties of the vector mesons (viz., $\rho$, $\omega$ and $\phi$ mesons)
produced in the inclusive photonuclear reaction at Jefferson Laboratory
(JLAB). The quoted mesons were detected by their dilepton, i.e., $e^+e^-$,
decay product.
The
electromagnetic probes, used by the CLAS collaboration, provide the
undistorted information about the properties of the vector meson in the
nucleus.

The CLAS Collaboration has analyzed the absorption of the $\omega$ and
$\phi$ mesons in the nucleus (to search the medium effect on these mesons)
by measuring the nuclear transparency ratios vs $A$ (mass number of the
nucleus) of the quoted mesons \cite{wood}.
It
should be mentioned that the transparency ratio depends on the elementary
vector meson nucleon total scattering cross section $\sigma^{*VN}_t$ (i.e.,
$\sigma^{*\omega N}_t$ and $\sigma^{*\phi N}_t$ for the considered mesons)
in the nucleus. The notation $V$ has been used to express either $\omega$ or
$\phi$.
The
data of the transparency ratio $T_A/T_C$ (i.e., normalized to $^{12}$C) for
the $\omega$ meson could not be reproduced by drastically increasing the
free space $\omega$ meson nucleon scattering cross section
$\sigma^{\omega N}_t$, i.e.,
$\sigma^{*\omega N}_t \geq 10\sigma^{\omega N}_t$ \cite{wood, dasA}.
On
the other hand, $T_A/T_C$ data for the $\phi$ meson reveal the drastic
enhancement of the $\phi$ meson nucleon scattering cross section in the
nucleus: $\sigma^{*\phi N}_t = (1.5-15) \sigma^{\phi N}_t$ \cite{wood, dasA}.
However,
the measured transparency ratios $T_A/T_d$ (i.e., normalized to deuteron)
for both $\omega$ and $\phi$ mesons can be explained by   
$\sigma^{*V N}_t \sim 2\sigma^{V N}_t$ \cite{dasA}.
It
should be mentioned that the elementary vector meson nucleon cross section
in the nucleus is related to the collision broadening of the vector meson,
i.e., $\Gamma^c_V \propto \sigma^{*VM}_t$.

The modification of the $\rho$ meson was observed by the CLAS collaboration
at JLAB \cite{nass} in the measured $e^+e^-$ invariant mass distribution
spectra in the photonuclear reaction. This collaboration has reported the
enhanced width (not mass-shift) of the $\rho$ meson photoproduced in the
nuclei.
The
calculated results for the quoted spectra also corroborate this finding
\cite{dasB, riek}. Being a broad resonance, the $\rho$ meson dominantly
decays inside the nucleus \cite{dasB}. Therefore, the invariant mass
distribution spectrum of its decay products (e.g., $\rho \to e^+e^-$) can
elucidate the medium modification of the $\rho$ meson.
In
contrast to it, the quoted spectrum of the decay products of the narrow
resonances (e.g., $\omega$ and $\phi$ mesons) cannot illustrate the in-medium
properties of these mesons as they dominantly decay outside the nucleus
\cite{dasL}.

The dilepton $e^+e^-$ emission due to the decay of the $\rho$ meson
coherently photoproduced in the nucleus had been studied earlier \cite{dasB},
where the nucleus remains in the same (ground) state after the reaction
occurred.
The
calculated $e^+e^-$ invariant mass distribution spectrum in the above
reaction shows the significant spreading (without mass-shift) of this meson,
specifically, in the heavy nucleus.
As
mentioned earlier, the CLAS collaboration \cite{nass} reported the in-medium
properties of the $\rho$ meson by measuring the $e^+e^-$ invariant mass
distribution spectra in the inclusive photonuclear reaction, where the final
state of the nucleus is not identified.
To
analyze this data, the cross sections of the above reaction on nuclei have
been calculated, and those are compared with the measured spectra
\cite{nass}.

\section{Formalism}

The dilepton $e^+e^-$ arises due to the decay of the $\rho$ meson produced
in the inclusive photonuclear reaction,
i.e.,
$ \gamma {\mbox A} \to \rho^0 {\mbox X} $; $\rho^0 \to e^+e^-$. The symbol
${\mbox A}$ represents the nucleus in the initial state, and
${\mbox X}$ is not specified in the final state.
The
$\rho$ meson (an unstable particle; $\tau_\rho \sim 10^{-23}$s) propagates
certain distance before it decays into dilepton: $\rho \to e^+e^-$. The
matrix element ${\cal M}_{fi}$ for the above reaction can be written as
\begin{equation}
{\cal M}_{fi} =
 \frac{e\gamma_{\gamma \rho}}{m^2}  
<\phi_{e^-}({\bf r}^\prime) \phi_{e^+}({\bf r}^\prime)  l_\mu
  G^{\mu\nu}_\rho ({\bf r}^\prime -{\bf r})  |R_{X0}({\bf r})|
 \epsilon_\nu ({\bf k}_\gamma, {\bf \lambda}_\gamma), \phi_\gamma ({\bf r})>,
\label{mtx}
\end{equation}
where $G^{\mu\nu}_\rho ({\bf r}^\prime -{\bf r})$ =
$(-g^{\mu \nu} +\frac{k^\mu_\rho k^\nu _\rho}{m^2})
G_\rho ({\bf r}^\prime -{\bf r})$ is the $\rho$ meson propagator. The second
part of it does not contribute to the cross section of the reaction. The
scalar part, i.e., $G_\rho ({\bf r}^\prime -{\bf r})$, describes the $\rho$
meson propagation from ${\bf r}$ to ${\bf r}^\prime$ in the space.
The
vector meson dominance (VMD) model \cite{saku, li} has been used to explain
the dielectron decay of the $\rho$ meson. $\gamma_{\gamma \rho}$ denotes the
$\rho$ meson to photon (virtual) conversion factor, as illustrated by VMD
model. In fact, it is connected to the $\rho$ meson hadron coupling constant
$f_\rho$ as
$\gamma_{\gamma \rho} =\frac{em^2_\rho}{f_\rho}$ \cite{saku, li}.
The
factor $1/m^2$ is the propagator of the virtual photon which coupled to
the leptonic current $l_\mu = {\bar u}_{e^-} \gamma_{\mu} v_{e^+}$ with the
coupling constant $e$ (electronic charge).
$\epsilon_\nu ({\bf k}_\gamma, {\bf \lambda}_\gamma)$ is the polarization
four vector of the photon, and $\phi$s are the wave functions for the
continuum particles.
$R_{X0}$
stands for the nuclear transition matrix element: $R_{X0}$({\bf r}) =
$\sum_j <X | {\bar f}_{\gamma N \to \rho N} ({\bf r} -{\bf r}_j) |A>$,
where
$A$ is the initial state (i.e., ground state) of the nucleus, and $X$
represents the unspecified final state.
$ {\bar f}_{\gamma N \to \rho N} ({\bf r} -{\bf r}_j) $
describes the production amplitude of the $\rho$ meson in the elementary
$\gamma N \to \rho N$ reaction occurring on the $j$th nucleon in the nucleus.

The matrix element ${\cal M}_{fi}$ can be decoupled as ${\cal M}_{fi}$
= $\Gamma (m_{s^-}, m_{s^+}, {\bf \lambda}_\gamma) F_\rho (X)$, where the
previous factor involves the internal coordinates of the continuum particles,
i.e.,
\begin{equation}
\Gamma (m_{s^-}, m_{s^+}, {\bf \lambda}_\gamma) = -
\frac{e\gamma_{\gamma \rho}}{m^2}
\bar {u} (k_{e^-}, m_{s^-}) \gamma_{\mu} v(k_{e^+}, m_{s^+})
\epsilon^{\mu} (k_\gamma, {\bf \lambda}_\gamma),
\label{geg}
\end{equation}
and
$F_\rho (X)$ illustrates the production, propagation and decay of the $\rho$
meson in the space:
\begin{equation}
F_\rho (X)
= \int \int d{\bf r} d{\bf r^\prime}
   e^{-i{\bf k}_\rho \cdot {\bf r^\prime}} G_\rho ({\bf r^\prime} -{\bf r})
   e^{i{\bf k}_\gamma \cdot {\bf r}} R_{X0} ({\bf r}),
\label{fxrh}
\end{equation}
with ${\bf k}_\rho (= {\bf k}_{e^+} + {\bf k}_{e^-})$ expressing the momentum
of the $\rho$ meson.

The form for $G_\rho ({\bf r^\prime} -{\bf r})$ \cite{golu, dasB} is given
by
\begin{equation}
G_\rho ({\bf r^\prime} -{\bf r})
= \delta ({\bf b^\prime} -{\bf b}) \theta (z^\prime -z)
   e^{i{\bf k}_\rho . ({\bf r^\prime} -{\bf r})}
   D_{{\bf k}_\rho} ({\bf b}, z^\prime, z),
\label{grh}
\end{equation}
where $D_{{\bf k}_\rho} ({\bf b}, z^\prime, z)$ comprises the interaction
of the $\rho$ meson with the nucleus:
\begin{equation}
D_{{\bf k}_\rho} ({\bf b}, z^\prime, z)
= -\frac{i}{2k_{\rho \parallel}} exp \left [ \frac{i}{2k_{\rho \parallel}}
   \int^{z^\prime}_z dz^{\prime \prime} \{ \tilde{G}^{-1}_{0\rho} (m)
   -2E_\rho V_{O\rho} ({\bf b}, z^{\prime \prime}) \} \right ].
\label{drh}
\end{equation}
In this equation, $V_{O\rho}$ symbolizes the $\rho$ meson optical potential
which modifies the parameters of this meson in the nucleus. $E_\rho$ is the
energy of the quoted meson.
$\tilde {G}^{-1}_{0\rho} [= m^2 -m^2_\rho +im_\rho\Gamma_\rho (m)]$
represents the inverse of the $\rho$ meson propagator of mass $m$ and total
decay width $\Gamma_\rho (m)$ in the free space. $m_\rho$(=775.26 MeV) is
the pole mass of this meson \cite{tana}.

Using $G_\rho ({\bf r^\prime} -{\bf r})$ given in Eq.~(\ref{grh}), the
expression of $F^{(X)}_\rho$ in Eq.~(\ref{fxrh}) can be simplified to
\begin{equation}
F_\rho (X)
= \int d{\bf r} e^{i{\bf q} \cdot {\bf r}} D_{k_\rho}({\bf b},z)
   R_{X0}({\bf r}),
\label{fxrh2}
\end{equation}
with ${\bf q}(={\bf k}_\gamma -{\bf k}_\rho)$ being the momentum transfer
to the nucleus. $D_{k_\rho}({\bf b},z)$ is given by
$D_{k_\rho}({\bf b},z)$ =
$ \int^\infty_z dz^\prime  D_{{\bf k}_\rho} ({\bf b}, z^\prime, z) $.

The differential cross section for the dilepton emission in the inclusive
photonuclear reaction can be written as
\begin{eqnarray}
d\sigma
= \frac{\pi^3}{(2\pi)^8} \frac{1}{E_\gamma E_{e^-} E_{e^+}}
  \delta^4(k_i-k_f) <|{\cal M}_{fi}|^2>
   d{\bf k}_{e^-} d{\bf k}_{e^+} d{\bf k}_X,
\label{dx1}
\end{eqnarray}
where $k_i(k_f)$ is the four momentum in the initial(final) state of the
reaction. $<|{\cal M}_{fi}|^2>$ refers to
\begin{eqnarray}
<|{\cal M}_{fi}|^2>
&=& \frac{1}{2} \sum_{{\bf \lambda}_\gamma} \sum_{m_{s^-}, m_{s^+}}
  \sum_X |{\cal M}_{fi}|^2    \nonumber  \\
&=& \frac{1}{2} \sum_{\lambda_\gamma} \sum_{m_{s^-}, m_{s^+}}
   |\Gamma (m_{s^-}, m_{s^+}, {\bf \lambda}_\gamma)|^2 \sum_X |F_\rho (X)|^2,   
\label{mxms}
\end{eqnarray}
where $\sum_X$ represents the summation over all nuclear final states.
${\bf \lambda}_\gamma$ is the polarization vector of the incoming photon.
$m_{s^-}$ and $m_{s^+}$ are the spin projections of the outgoing electron
and positron respectively.

Using Eq.~(\ref{mxms}), the differential cross section for the dilepton
$e^+e^-$ invariant mass (i.e., $\rho$ meson mass) distribution in the
reaction quoted in Eq.~(\ref{dx1}) can be expressed as
\begin{equation}
\frac{d\sigma (E_\gamma)}{dm d\Omega_\rho} = K_F m^2
\Gamma(m)_{\rho^0 \to e^+e^-} |{\bar f}(0)_{\gamma N \to \rho N}|^2
\int d{\bf r} |D_{{\bf k}_\rho} ({\bf b},z)|^2 \varrho ({\bf r}),
\label{dx2}
\end{equation}
with $K_F$ = $\frac{3\pi}{(2\pi)^4}$$\frac{k^2_\rho (E_i-E_\rho)}
{E_\gamma |k_\rho E_i - {\bf k}_\gamma \cdot {\hat k}_\rho E_\rho|}$.
$\Gamma(m)_{\rho^0 \to e^+e^-}$ is the width of the $\rho$ meson of mass
$m$ decaying at rest into $e^+e^-$.

The differential cross section stated in Eq.~(\ref{dx2}) is  based on the
fixed scatterer or frozen nucleon (in the nucleus) approximation. The
Fermi motion of the nucleon in the nucleus can be incorporated replacing
${\bar f}_{\gamma N \to \rho N}$ by
$< {\bar f}_{\gamma N \to \rho N} >_A$, i.e.,
\begin{equation}
< {\bar f}(0)_{\gamma N \to \rho N} >_A
= \int \int d{\bf k}_N d\epsilon_N S_A ({\bf k}_N, \epsilon_N) 
  {\bar f}(0,s)_{\gamma N \to \rho N},
\label{gNN}
\end{equation}
with
\begin{eqnarray}
s &=& (E_\gamma+E_N)^2 - ({\bf k}_\gamma + {\bf k}_N)^2;   \nonumber  \\
E_N &=& m_A - \sqrt{ k_N^2+(m_A -m_N +\epsilon_N)^2 },   \nonumber
\label{sEN}
\end{eqnarray}
where $S_A ({\bf k}_N, \epsilon_N)$ represents the spectral function of
the nucleus, normalized to unity. It illustrates the probability of a
nucleon with momentum ${\bf k}_N$ and binding energy $\epsilon_N$ in the
nucleus \cite{pary}. $S_A ({\bf k}_N, \epsilon_N)$ for various nuclei are
discussed elaborately in Ref.~\cite{pary2}. Therefore, those have not been
presented explicitly.

It should be mentioned that the tagged photon beam was used for the
experiment done by the CLAS collaboration at Jlab \cite{nass}. The quoted
beam possesses certain energy range which can be weighted in 6 bins to
simulate the beam profile \cite{riek} (also see the references there in).
Therefore, the cross section of the considered reaction due to tagged
$\gamma$ beam can be written \cite{dasB} as
\begin{equation}
\frac{ d\sigma }{ dm } = \sum^6_{i=1} W(E_{\gamma, i})
          \frac{ d\sigma(E_{\gamma, i}) }{ dm },
\label{dxtb}
\end{equation}
where $W(E_{\gamma, i})$ is the relative weights of $13.7\%$, $23.5\%$,
$19.3\%$, $20.1\%$, $12.6\%$ and $10.9\%$ for $E_{\gamma, i}$(GeV) equal
to 1.0, 1.5, 2.0, 2.5, 3.0 and 3.5 respectively \cite{riek}.

\section{Result and Discussions}

The total decay width $\Gamma_\rho (m)$ of the $\rho$ meson in the free
space is composed of the partial widths of the $\rho$ meson decaying into
various channels, i.e., $\Gamma_\rho (m) \approx$
$99.94 \times 10^{-2} \Gamma (m)_{\rho \to \pi^+\pi^-}$
+ $6 \times 10^{-4} \Gamma (m)_{\rho \to \pi^0\gamma}$ \cite{tana}. The
two body decay width of the $\rho$ meson $\Gamma (m)_{\rho \to d_1d_2}$
\cite{manl} is given by
\begin{equation}
\Gamma (m)_{\rho \to d_1d_2} = \Gamma (m_\rho)_{\rho \to d_1d_2}
\left [ \frac{\Phi_l(m)}{\Phi_l(m_\rho)} \right ].
\label{pdw}
\end{equation}
The values of $ \Gamma (m_\rho)_{\rho \to \pi^+\pi^-} $ and
$ \Gamma (m_\rho)_{\rho \to \pi^0\gamma} $ are tabulated in Ref.~\cite{tana}.
$\Phi_l$ represents the phase space factor of the two body decay:
$ \Phi_l(m) =\frac{ {\tilde k} }{ m } B^2_l({\tilde k}R) $, with the
interaction radius $R=1$ fm \cite{effen}. ${\tilde k}$ is the momentum in
the cm system of the decay products of the $\rho$ meson.
The
angular momentum $l$ associated with the considered $\rho$ meson decay
channels is equal to unity. $B^2_l({\tilde k}R)$ represents the
Blatt-Weisskopf barrier penetration factor:
$B^2_1(\mbox{x}) = \frac{\mbox{x}^2}{1+\mbox{x}^2}$ \cite{manl}.
The
dielectron decay width of the $\rho$ meson, negligibly small compared to
$\Gamma_\rho (m)$, used in Eq.~(\ref{dx2}) is given by
$\Gamma(m)_{\rho^0 \to e^+e^-}$ =
$\frac{\pi}{3} (\frac{\alpha_{em}}{\gamma_\rho})^2 \frac{m^4_\rho}{m^3}$
\cite{li}. $\alpha_{em} (=1/137.04)$ denotes the fine structure constant,
and $\gamma_\rho$ is equal to half of the $\rho$ meson hadron coupling
constant $f_\rho (=2\gamma_\rho)$, as mentioned earlier. The value of
$\gamma_\rho$ (= 2.48 \cite{dasS}) is directly determined from the measured
$\Gamma(m_\rho)_{\rho \to e^+e^-}$ \cite{tana}.

The $\rho$ meson production amplitude for the elementary
$\gamma N \to \rho N$ reaction can be expressed as
${\bar f}_{\gamma N \to \rho N}$
= -$ 4\pi E_\rho [\frac{1}{{\tilde E}_\rho} + \frac{1}{{\tilde E}_N}]
{\tilde f}_{\gamma N \to \rho N} $ \cite{dasH}.
The
symbol ``$\tilde {~}$'' on the quantities represents those assessed at
$\gamma N$ cm energy. $f_{\gamma N \to \rho N}$ denotes the reaction
amplitude of the quoted elementary reaction.
The
vector meson dominance model connects $f_{\gamma N \to \rho N}$ to the
$\rho N \to \rho N$ scattering amplitude $f_{\rho N \to \rho N}$ as
$ f_{\gamma N \to \rho N} $
=$ \frac{\sqrt{\pi \alpha_{em}}}{\gamma_\rho} f_{\rho N \to \rho N} $
\cite{dasS, bauer}. The energy dependent experimentally determined values
of the forward $f_{\rho N \to \rho N}$ are given in Ref.~\cite{kond}.

The optical potential of the $\rho$ meson $V_{O\rho}$, which describes the
$\rho$ meson nucleus interaction, can be expressed \cite{glau} as
\begin{equation}
V_{O\rho} ({\bf r}) 
= -\frac{v_\rho}{2} (\alpha_{\rho N}+i) \sigma^{\rho N}_t \varrho ({\bf r}),
\label{Opf}
\end{equation}
where
$v_\rho$ is the velocity of the $\rho$ meson. $\alpha_{\rho N}$ represents
the ratio of the real to imaginary part of $f_{\rho N \to \rho N} (0)$ in
the free space.
$\sigma^{\rho N}_t$
denotes the total $\rho$ meson nucleon scattering cross section:
$\sigma^{\rho N}_t$ = $\frac{4\pi}{k_\rho} f_{\rho N \to \rho N} (0)$.
$\varrho ({\bf r})$
symbolizes the density distribution of the nucleus, normalized to the mass
number of the nucleus. The form for it, as extracted from the electron
scattering data \cite{jage}, is used to evaluate $V_{O\rho} ({\bf r})$.

\begin{figure}
\begin{center}
\centerline {\vbox {
%\psdraft
\psfig{figure=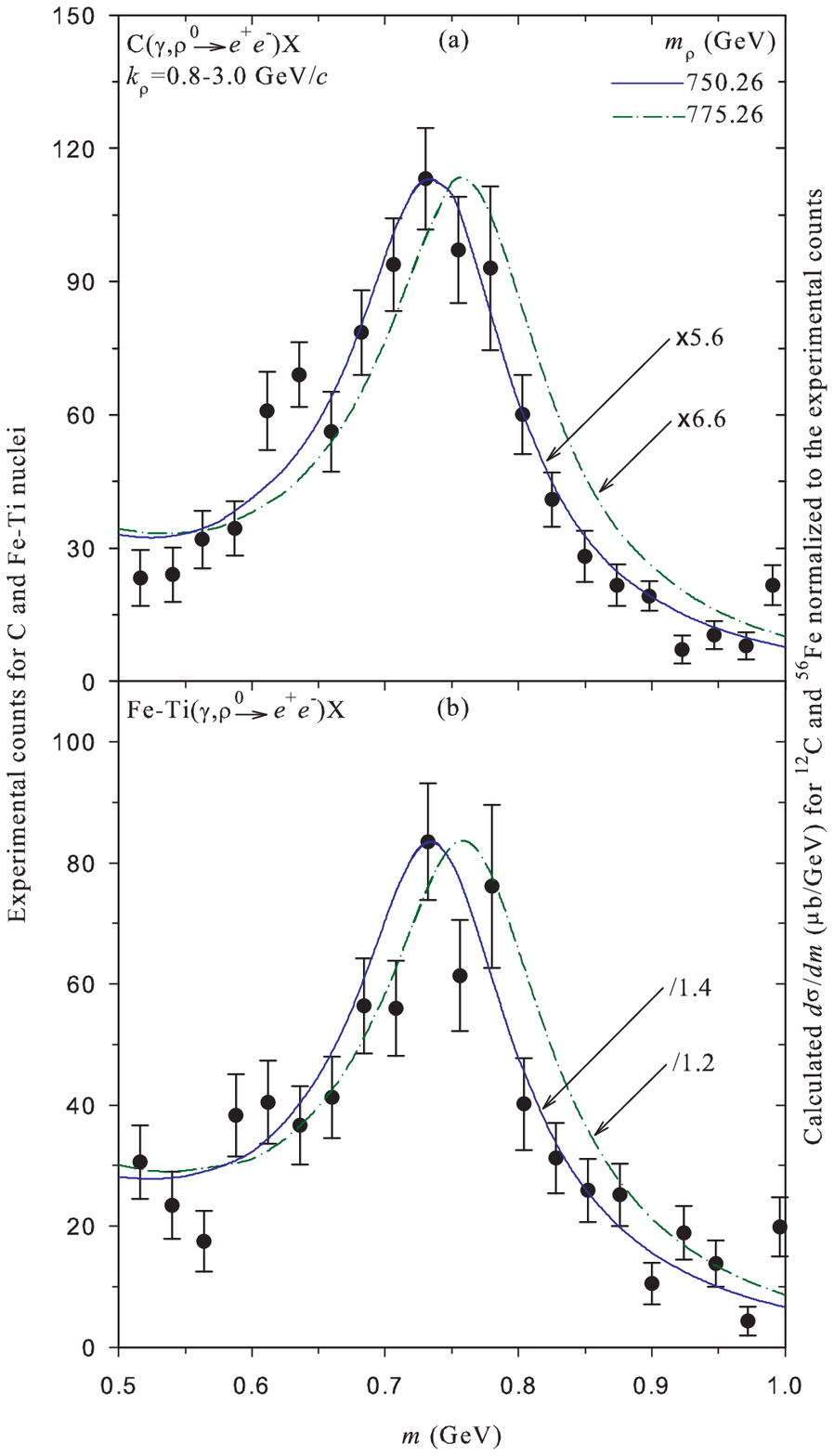,height=14.0 cm,width=08.0 cm}
}}
\caption{
(color online).
The differential cross sections $\frac{d\sigma}{dm}$ calculated for the
$\rho$ meson mass distribution in the inclusive photoinduced reactions on
$^{12}$C and $^{56}$Fe are compared with data taken for C and Fe-Ti nuclei
\cite{nass} respectively. The calculated results are normalized to the
experimental counts.
}
\label{Fgdt}
\end{center}
\end{figure}

The differential cross sections $\frac{d\sigma}{dm}$ of the $\rho$ meson
mass $m$ (i.e., $e^+e^-$ invariant mass) distribution in the inclusive
photonuclear reaction, i.e., A$(\gamma, \rho \to e^+e^-)$X, have been
calculated for the $\rho$ meson momentum $k_\rho$ = 0.8-3.0 GeV/$c$, as that
was the restriction on $k_\rho$ imposed in the CLAS measurements \cite{nass}.
The
calculated $\frac{d\sigma}{dm}$ for $^{12}$C and $^{56}$Fe nuclei (normalized
to the experimental counts for C and Fe-Ti nuclei \cite{nass} respectively)
have been presented in Fig.~\ref{Fgdt} for the $\rho$ meson pole mass
$m_\rho$ taken equal to 750.26 MeV (solid curve) and 775.26 MeV (dot-dashed
curve).
As
visible in this figure, the reduction of $m_\rho$ from 775.26 MeV to 750.26
MeV ($\sim 3 \%$) shows good agreement of the calculated results with the
measured spectra for all nuclei. It is remarkable that the quoted reduction
of $m_\rho$ is consistence with the mass-shift parameter ($\alpha_\rho =
0.02 \pm 0.02$) reported by the CLAS collaboration \cite{nass}. Therefore,
$m_\rho$ is taken equal to 750.26 MeV to investigate the modification of the
$\rho$ meson parameters in the nucleus.

\begin{figure}
\begin{center}
\centerline {\vbox {
%\psdraft
\psfig{figure=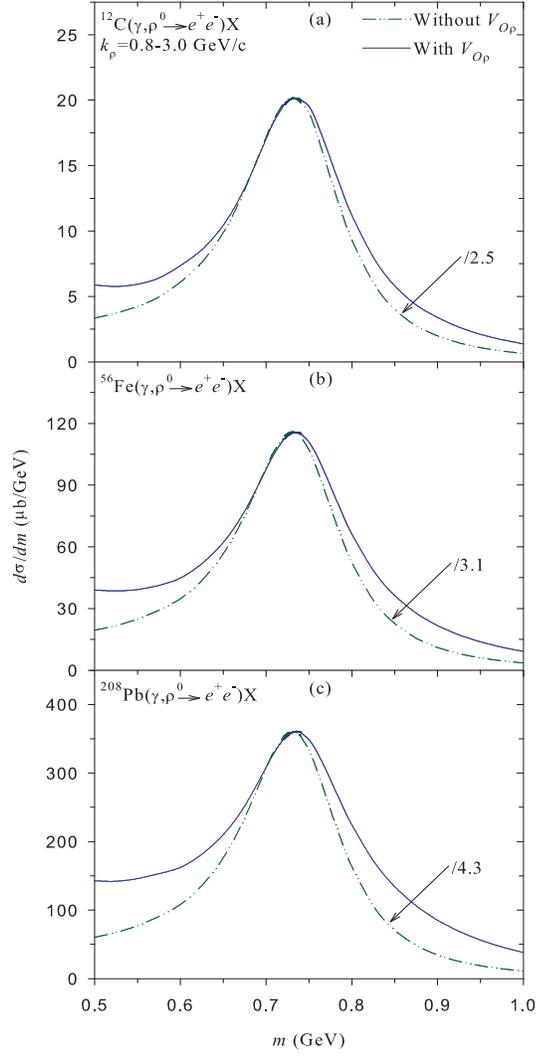,height=14.0 cm,width=07.0 cm}
}}
\caption{
(color online).
The $\rho$ meson mass distribution spectra calculated with and without the
$\rho$ meson optical potential $V_{O\rho}$ are presented for nuclei. The
broadening in the width (not peak-shift) of the spectrum due to $V_{O\rho}$
is distinctly visible in the figure.
}
\label{Fgmd}
\end{center}
\end{figure}

The cross sections $\frac{d\sigma}{dm}$ with and without incorporating the
$\rho$ meson optical potential $V_{O\rho}$ are calculated to disentangle the
nuclear effects on the $\rho$ meson.
The
calculated results for $^{12}$C, $^{56}$Fe and $^{208}$Pb nuclei are depicted
in Fig.~\ref{Fgmd}. The dot-dot-dashed curve illustrates $\frac{d\sigma}{dm}$
assessed without $V_{O\rho}$ and the solid curve represents that evaluated
including $V_{O\rho}$ in the calculation.
This
figure distinctly elucidates the broadening in the width of the $\rho$ meson
mass distribution spectra due to $V_{O\rho}$ for all nuclei without the
shift in the peak position, i.e., mass-shift.
The
quantitative values of the widths of the $\rho$ meson mass distribution
spectra $\Gamma^*_\rho$, and the enhancement in the widths
$\Delta \Gamma_\rho$ due to $V_{O\rho}$ for the above mentioned nuclei are
listed in table 1.

\begin{figure}
\begin{center}
\centerline {\vbox {
%\psdraft
\psfig{figure=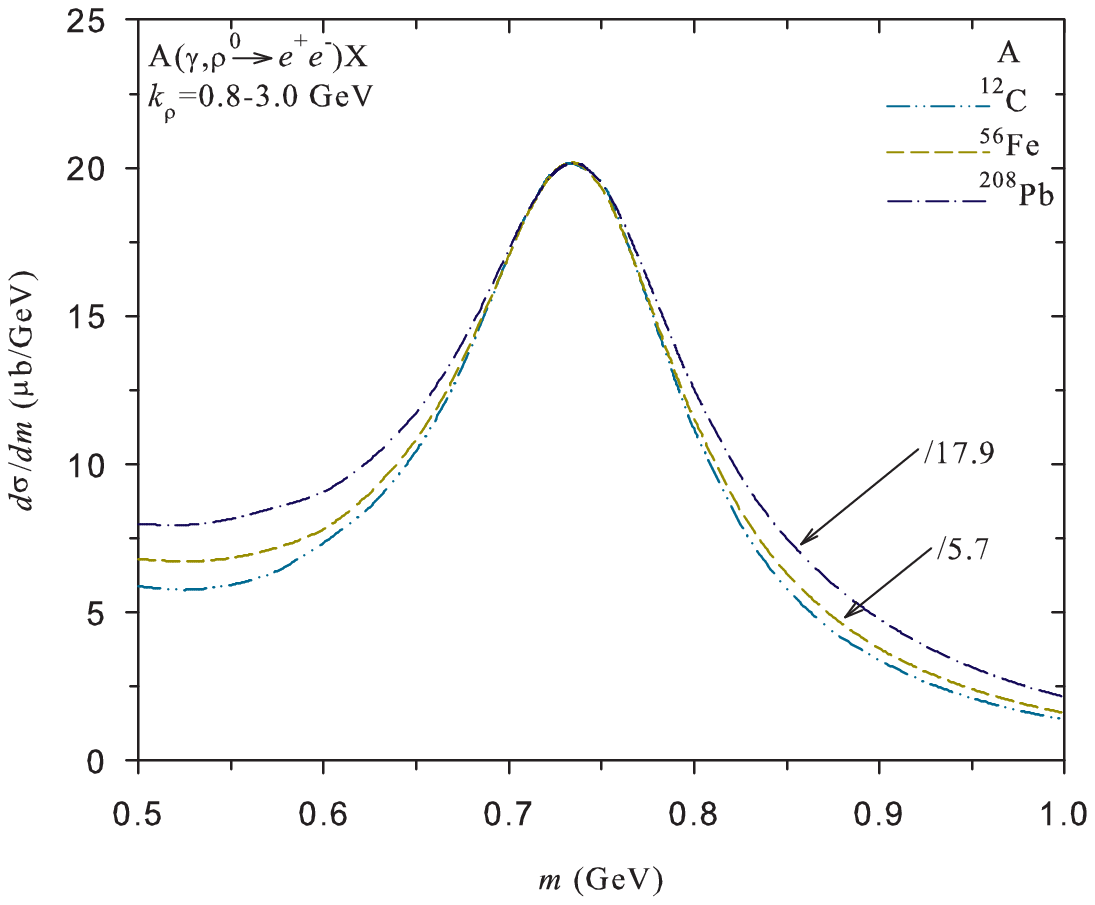,height=08.0 cm,width=08.0 cm}
}}
\caption{
(color online).
The $\rho$ meson mass distribution spectra for $^{12}$C, $^{56}$Fe, and
$^{208}$Pb nuclei are compared. The enhancement in the width of the spectrum
due to $V_{O\rho}$ increases with the size of the nucleus.
}
\label{FgmA}
\end{center}
\end{figure}

The calculated $\frac{d\sigma}{dm}$ are compared in Fig.~\ref{FgmA} for
$^{12}$C (dot-dot-dashed curve), $^{56}$Fe (dashed curve) and $^{208}$Pb
(dot-dashed curve) nuclei to visualize the relative broadening in the $\rho$
meson mass distribution spectra.
It
is noticeable in the figure that the spreading of the spectrum increases
with the size of the nucleus. Therefore, the medium modification of the
$\rho$ meson properties can be demonstrated better in the heavy nucleus.

\begin{table}[ht]
%\begin{table}
\caption{The calculated width of the $\rho$ meson mass distribution spectrum
$\Gamma^*_\rho$, and the enhancement in the width $\Delta \Gamma_\rho$
due to the $\rho$ meson nucleus interaction $V_{O\rho}$.}
\centering
\begin{tabular}{c c c}
\hline
  Nucleus   & $\Gamma^*_\rho$ (MeV) & $\Delta \Gamma_\rho$ (MeV) \\
\hline
 $^{12}$C   &  163.5                &    19.6                    \\
 $^{56}$Fe  &  172.1                &    29.4                    \\
 $^{208}$Pb &  197.5                &    54.8                    \\
\hline
\end{tabular}
\label{tab}
\end{table}

\section{Conclusions}

The differential cross section for the $\rho$ meson mass distribution in the
inclusive photonuclear reaction has been calculated to investigate the
modification of the hadronic parameters (i.e., mass and width) of the $\rho$
meson in the nucleus.
The
quoted modification arises due to the interaction of the $\rho$ meson with
the nucleus, which is described by the $\rho$ meson optical potential. It
is evaluated by folding the $\rho$ meson nucleon scattering amplitude with
the density distribution of the nucleus.
The
calculated result does not show the mass modification in the $\rho$ meson
mass distribution spectrum, but that illustrates the increase in the width
of the quoted spectrum. The enhancement in the width increases with the mass
of the nucleus.
The
calculated $\rho$ meson mass distribution spectra for nuclei reproduce well
the data reported from the Jefferson Laboratory.

\section{Acknowledgement}

The author appreciates Dr. S. M. Yusuf for the support and encouragement to
work on theoretical nuclear physics.

\newpage
%\vspace{1 cm}
%\begin{figure}[h]

\newpage
%\vspace{1 cm}
%\begin{figure}[h]

\newpage
%\vspace{1 cm}
%\begin{figure}[h]

\newpage

{\bf Figure Captions}
\begin{enumerate}

\item
(color online).
The differential cross sections $\frac{d\sigma}{dm}$ calculated for the
$\rho$ meson mass distribution in the inclusive photoinduced reactions on
$^{12}$C and $^{56}$Fe are compared with data taken for C and Fe-Ti nuclei
\cite{nass} respectively. The calculated results are normalized to the
experimental counts.

\item
(color online).
The $\rho$ meson mass distribution spectra calculated with and without the
$\rho$ meson optical potential $V_{O\rho}$ are presented for nuclei. The
broadening in the width (not peak-shift) of the spectrum due to $V_{O\rho}$
is distinctly visible in the figure.

\item
(color online).
The $\rho$ meson mass distribution spectra for $^{12}$C, $^{56}$Fe, and
$^{208}$Pb nuclei are compared. The enhancement in the width of the spectrum
due to $V_{O\rho}$ increases with the size of the nucleus.

\end{enumerate}

\end{document}